\documentclass[prd,a4paper,twocolumn]{revtex4} 

\usepackage{graphicx} 
\usepackage{subfigure,amssymb,latexsym} 
\usepackage{psfrag} 

\newcommand{\nc}{\newcommand} 

\nc{\be}[1]{\begin{equation}\mbox{$\label{#1}$}} 
\nc{\bea}[1]{\begin{eqnarray} \mbox{$\label{#1}$}} 
\nc{\eea}{\end{eqnarray}} 
\nc{\ee}{\end{equation}} 

\nc{\bc}{\begin{center}} 
\nc{\ec}{\end{center}} 
\nc{\ba}{\begin{array}} 
\nc{\ea}{\end{array}} 
\nc{\bab}{\begin{abstract}} 
\nc{\eab}{\end{abstract}} 
\nc{\btab}{\begin{tabular}} 
\nc{\etab}{\end{tabular}} 
\nc{\bit}{\begin{itemize}} 
\nc{\eit}{\end{itemize}} 
\nc{\ben}{\begin{enumerate}} 
\nc{\een}{\end{enumerate}} 
\nc{\bfig}{\begin{figure}} 
\nc{\efig}{\end{figure}} 

\nc{\dd}[2]{{{\partial #1}\over{\partial #2}}} 
\nc{\ddd}[2]{{{{\partial}^2 #1}\over{\partial {#2}^2}}} 
\nc{\dddd}[3]{{{{\partial}^2 #1}\over 
   {\partial #2 \partial #3}}} 

\nc{\ie}{{\it i.e. }} 
\nc{\eg}{{\it e.g. }} 

\nc{\mn}{{\mu\nu}} 

\nc{\gppn}{$\gamma_{PPN} \,$} 
\nc{\lcdm}{$\Lambda$CDM\,} 
\nc{\f}{\frac}

\begin{document} 


\title{Consistency of f(R) gravity models around solar polytropes}

\author{K. Henttunen} 
\email{kajohe@utu.fi} 
\author{I. Vilja} 
\email{vilja@utu.fi} 
\affiliation{Department of Physics and Astronomy, University of Turku, FIN-20014 Turku, FINLAND} 

\date{\today} 

\begin{abstract} 
It is stated that a class of $f(R)$ gravity models seem to obtain \lcdm  
behaviour for high redshifts and general relativistic behaviour 
locally at high curvatures. In the present paper, we numerically study polytropic
configurations that resemble stars like young sun with Hu and Sawicki
$f(R)$ gravity field equations
and compare the spacetime at the boundary to the general relativistic 
counterpart. 
These polytropes are stationary spherically symmetric configurations 
and have regular metrics at the origin.
Since Birkhoff's theorem does not apply for modified gravity, the solution
outside may deviate from Schwarzschild-de Sitter spacetime.
At the boundary, Post-Newtonian parametrization was used
to determine how much the studied model deviates from the general relativistic 
\lcdm model.

\end{abstract} 

\maketitle 

\section{Introduction}
According to observations \cite{sn1a,cmb}, general relativity alone cannot explain
the apparent acceleration of the Universe.  
An invisible and unclustered  dark energy (DE) component  is needed to 
confront the observations.  
In the simplest modification of Einstein's general relativity (GR), 
the \lcdm model (see \eg \cite{lcdm} and references therein),
Einstein-Hilbert action is modified by a tiny constant $\Lambda$. 
With this deviation, the universe accelerates according to 
cosmic background supernovae and large scale observations 
\cite{cmb,sn1a,bao,isw}.
This model, however, has several profound shortcomings considering \eg its tiny value and
the coincidence problem \cite{coincid}. There has not been a satisfactory explanation for these
problems, so other theories that could explain the observations have been sought for.
Various other means to introduce the accelerated expansion into the field equations  
include \eg cosmological scalar fields \cite{quint}, vector fields \cite{teves},  
string theory  and higher dimension inspired models \cite{string-dim} 
as well as several type of gravity modifications \cite{etg}.

Modifications to  gravity include models where the action function consists
 of the Ricci scalar and other curvature invariants 
($f(R,R_{\mn}R^{\mn},R_{\mn\alpha\beta}R^{\mn\alpha\beta},G,...)$, where the Gauss-Bonnet invariant is
$G=R^2-4R_{\mn}R^{\mn}+R_{\mn\alpha\beta}R^{\mn\alpha\beta}$).
Also, recently there has been much interest in so called General Galileon models, that describe 
the most general covariant scalar field theory coupled to gravity.

Simplest class of these are the
$f(R)$ gravity theories \cite{fR_intro,tsujikawa}, 
where the Einstein-Hilbert action is 
generalized with a function of the Ricci scalar only. These   
provide natural alternatives for dark energy  that
needs new physics to explain the accelerated expansion of the universe,
atural in the sense that no new exotic fields need to be included
on top of those already observed.
However, many $f(R)$ functions seem to act equivalently to Brans-Dicke
scalar fields. This makes possible to treat the model in different frameworks for 
consistency check.

In a successful gravity modification, subdominant terms in the action functional become  
essential at low curvature and hence accelerate the expansion of the universe. 
The observed large scale defects could then be explained by 
the deviation from Einstein-Hilbert gravity. However,  Einstein-Hilbert gravity 
is well tested within solar system length scales \cite{GRtests}. 
Therefore, we focus on models considered as    
alternatives for the concordance \lcdm  model with GR  like 
behaviour at solar system scale. 

When the modification in the action functional is written using 
auxiliary scalar fields (as in quintessence and scalar-tensor theories \cite{quint,etg}), 
the new effects of gravity can be interpreted as scalar-field - matter interactions. 
Chameleon $f(R)$ models \cite{chameleon} allow the mass of the scalar  
field to vary according to local matter density and 
give rise to non-linear self-interactions.  
With chameleon behaviour, correction to the Newtonian gravitational  
potential may be small in high curvature regimes (\eg near the sun) and 
the metric therefore close to Schwarzschild-de Sitter (SdS) solution \cite{MTW}.

While the metric outside a spherically symmetric body in GR must be the  
Schwarzchild-de Sitter spacetime according to Birkhoff's theorem \cite{WEI},  
this does not hold for a spherically symmetric matter distributions in non-trivial  
$f(R)$ theories. In these theories, the way spacetime 
outside the mass distribution 
depends on  the solution inside is more complex \cite{MV,birk1}.

There are many $f(R)$ models that can produce correct shift from matter  
domination to radiation domination and to the  
currently accelerating epoch (see \eg \cite{AT_mr,LinGuChen}).  
Also, constraints from structure formation and stability conditions  
are satisfied by various models \cite{Seifert,dolgov,woodard,clifton,faraoni}. 
Since gravity is most rigorously tested at local scales and with low  
energies tangible $f(R)$ model must be able to settle to the observed   
GR values within this regime. 
The $f(R)$ models that pass all the abovementioned tests  are scarce.
Among current viable $f(R)$ models are
\cite{HS,AB,St,NO1,NO2}.

In the present paper, we study the Hu and Sawicki $f(R)$-function (HS) numerically. 
Observations of external gravity provide constraints for the internal stellar structure.
Therefore,
we have compared static spherically symmetric bodies with a polytropic equation  
of state with the HS $f(R)$ and Einsteinian gravity. 
These constraints are studied in the present paper by finding out how widely initial
conditions at the center lead to an acceptable spacetime at the boundary. While Hu and Sawicki
argue to satisfy the solar system constraints with the thin-shell condition by
considering external fields in the solar vicinity. We study the stellar interior only
and find out numerically for a wide model parameter range
that, HS-model follows closely the general relativistic
high curvature profile almost throughout the stellar interior, but does not
always arrive at the GR value at the boundary.
We particulary  want to find out how well these models correspond to GR 
around sun like polytropes by comparing a spacetime parameter outside to the
Cassini observations \cite{cassini}.

We also check the mass of the auxiliary scalar degree of freedom for the 
used equaton ansatz and  
find it to be high enough to evade the ``fifth force'' problem 
\cite{chameleon}.  

This paper is organized as follows: first in Section \ref{eqns} the gravitational 
framework is discussed, in Section \ref{sphsymm} the polytropic configuration and the 
surrounding spacetime is constrained and 
 in Section \ref{numerical} the numerical work is represented.
Finally in Section \ref{conclu} we discuss the results.

\section{Gravity formalism}\label{eqns}

We consider $f(R)$ gravity action \cite{HS} of the form   
\be{action} 
S = \int{d^4x\,\sqrt{-g}\Big(\frac{1}{2\kappa^2}f(R)+{\cal{L}}_{m}\Big)}, 
\ee 
where $\kappa^2=8\pi G$, and ${\cal{L}}_{m}$ is the matter Lagrangian. 
The corresponding field equations, derived in the metric formalism, are 
\be{eequs} 
F(R) R_{\mu\nu}-\frac 12 f(R) g_{\mu\nu}-(\nabla_\mu\nabla_\nu -g_{\mu\nu}\Box ) 
F(R)=8 \pi G T_{\mu\nu}. 
\ee 
Here $T_{\mu\nu}$ is the standard minimally coupled stress-energy tensor 
for perfect fluid and 
$F(R)\equiv df/dR$.  
By contracting the field equations we get the trace equation 
\be{traceq} 
F(R)R - 2 f(R) + 3\Box F(R) = 8 \pi G (\rho - 3 p). 
\ee 
which can be used as one independent dynamical equation. 

As usual, the matter term needs to obey the equation of continuity  
$D_\mu T^{\mu\nu}=0$. 
The only non-trivial component for spherically symmetric system  
here is 
\be{cont} 
p'=-\frac{B'}{2 B}(\rho+p), 
\ee 
where comma stands for radial derivative, $'\equiv d/dr$.  

Like in GR, the equation of continuity is  
automatically satisfied here if the field equations are satisfied  
\cite{koivisto} and no additional information is gained. 
The calculations were highly dependent on the equation ansatz.
Therefore, we chose the continuity  
equation to be one of the solved dynamical equations   
instead of using the full set of field equations.  

As the set of independent equations we use  the \linebreak  
$1 1$ -component (corresponding to $T_{11}$)  of modified field equations (\ref{eequs}),  
the trace equation (\ref{traceq}), the definition for the Ricci scalar in 
terms of metric  
components (\ref{metric})  and the continuity equation (\ref{cont}).

\subsection{The Hu-Sawicki  $f(R)$ model}\label{HSchapt}

There are still several physically attractable  modified gravity models 
among metric $f(R)$ gravity models
\cite{faraoni,comp-fRs} that can describe the observed expansion of the universe.
Hu-Sawicki $f(R)$ model vanishes for flat spacetime, $f(R)=0$ as $R\rightarrow 0$
and is able to reproduce  
$\Lambda CDM$ behavior as a limiting case as $R\rightarrow \infty$.
Correct cosmological dynamics as well as smooth transitions between different eras
are expected as the modifications are effective only at low-redshifts.
The Ostrogradski and matter instabilities are also avoided
\cite{HS,Seifert,dolgov,woodard}. Putting all  this together,
Hu-Sawicki model seems to be a good rival for \lcdm.

In the present paper, we compare this non-linear metric $f(R)$ model  to
GR in the weak-field regime. General relativistic behavior is
expected to emerge in higher curvature than the cosmic background.
Differences are found around a static spherically symmetric matter density with a 
polytropic equation of state. 

According to \cite{HS}, the Hu and Sawicki model exhibits the cameleon mechanism.
In this scenario,
the exterior gravitational field is generated by a thin shell of mass that lies about 
a thin transition region from the high curvature region 
to a low galactic background value.
Steep enough change in the potential of the scalar field
prevents the interior field to be detected.
They conclude that the model should correctly arrive at the 
solar system observational value outside a sun like matter distribution with a sufficiently high
galactic value for the field $f_{Rg}$.

Hu and Sawicki state a condition for the high curvature solution to occur
when the field gradients can be ignored wrt the source density, 
either throughout the whole system or locally, in the latter case the interior 
solution depends on the exterior solution.
When this property is applicable, the scalar degree of freedom is locked to the 
general relativistic value  and the exterior field is generated
only by the thin shell of mass outside the transition region between the interior 
and exterior field values.
This occurs when their thin-shell criterion is 
first satisfied going from outside in \cite{HS}. 
They calculated (\gppn $ -1$) to peak far from the star (around Jupiter's distance at $r=1000\, r_{/odot}$) 
at a value of the order $\sim 10^{-15}$.

In this work we calculate the Parameterized Post-Newtonian (PPN) parameter \cite{MTW}
\gppn at the surface by solving a
static spherically symmetric polytrope with non-linear Hu-Sawicki modified field equations.
We solve for the polytropes only within the solar interior
and calculate the \gppn value near the photosphere where
it was measured by Cassini mission.
We do not resort to the thin-shell mechanism in this work.
The scalar curvature inside the polytrope in our calculations  follows
the general relativistic high curvature value $R\sim\kappa \rho$ well
inside the stellar interior for all the tested model parameter values.

The $f(R)$ function for the Hu-Sawicki model  is given by \cite{HS} 
\be{fHS} 
f(R)_{HS}=R-m^2\frac{c_1(\frac{R}{m^2})^n}{c_2(\frac{R}{m^2})^n+1}.
\ee 
This contains three essential parameters, positive real numbers $c_1,\ c_2$ and a 
positive integer $n$. The parameter $m$ can be always included in parameters  
$c_1,\ c_2$, but it is convenient to use it as the mass 
scale chosen in the paper \cite{HS},  
\be{m-HS} 
m^2\equiv \frac{\kappa^2\rho_0}{3} = (8315 \textrm{Mpc})^{-2}  
\bigg(\frac{\Omega_m h^2}{0.13}\bigg). 
\ee 
Here $\rho_0$ is the average energy density today and $h$ is the reduced Hubble parameter.
In our work $h=0.72$.

One useful way to study this family of $f(R)$ models is to write (\ref{fHS}) as: 
\be{modfHS} 
f(R)=R-\f{c_1 m^2}{c_2}+\f{c_1 m^2}{c_2\left(1+ c_2\left(\f{R}{m^2}\right)^n  
\right)}. 
\ee 
Now one finds that only when the Ricci scalar $R$ is near the vacuum value 
$m^2$, can the third term contribute to \lcdm. However, because of the  
non-linear nature of the field equations, behaviour of the solution
may be different from \lcdm.  

The authors of \cite{HS} expand the function $f(R)_{HS}-R$, for $R>>m^2$
and arrive at \lcdm cosmology with $c_1/c_2^2\rightarrow 0$.
The function (\ref{fHS}) is expanded at curvatures that are high compared with (\ref{m-HS}).
We compare this model to \lcdm  by using the
full non-linear equations of motion that allow the Birkhoff's  
theorem to be violated and the \gppn parameter to settle down to a different
value than the SdS \gppn$=1$ near the stellar surface.

\section{Static spherically symmetric solutions} \label{sphsymm}
We compare the Hu-Sawicki gravity model with a time-independent 
spherically 
symmetric metric to the general relativistic case, where a non-rotating 
spherically 
symmetric matter configuration produces the Schwarzschild-de Sitter (SdS) 
solution near the object. 

The analysis is carried out by
adopting the following spherically symmetric, static, isotropic line element
((8.1.4) in \cite{WEI}) for the
spacetime configuration
\be{metric} 
ds^2=B(r)dt^2-A(r)dr^2-r^2(d\Theta^2-\sin^2\Theta\, d\Phi^2). 
\ee 
Comparing to the SdS solution, which is the spherically symmetric 
vacuum solution of Einstein's field equations, 
the metric components read
\bea{schw-metric}
B(r)=c^2 \left(1- \f{2 G m}{c^2 r} - \f{\Lambda}{3} r^2\right),\\
A(r)=\left(1- \f{2 G m}{c^2 r} - \f{\Lambda}{3} r^2\right)^{-1}.
\eea
This solution is also a good approximation for the exterior of a 
slowly rotating body like the sun.

\subsection{Spacetime constraints} 

If the studied $f(R)$ modification (\ref{fHS}) is to produce 
general relativistic gravitation within the weak field of solar system, the 
Schwarzschild-de Sitter spacetime is anticipated as the resulting spacetime  
around the calculated matter distributions. 
We calculate the  Post-Newtonian \gppn parameter at the boundary 
to see how well the general relativistic  
value, \gppn$=1$, is reproduced. 

The Post-Newtonian parametrization gives the strongest bound for a  
modified gravity theory within the solar system. 
Observational constraints are presented in Table \ref{PPNs}.  
The  Cassini mission \cite{cassini,Will} measurement 
constrains the spacetime parameter \gppn most stringently within the solar system: 
\be{cassinig} 
\gamma_{PPN}-1= (2.1\pm2.3)\times 10^{-5}. 
\ee 
Therefore, this Cassini constraint is used as a reference for choosing the 
acceptable  solutions that mimic general relativistic \lcdm polytropes. 

Derivation of the used \gppn parameter
was done in \cite{HMV}, 
and the parameter is expressed as
\be{gamppn} 
\gamma_{PPN}=\f{1+\f{B(r)}{r B'(r)}}{1-\f{A(r)}{r A'(r)}}.
\ee 
In this formula, the cosmological term $\Lambda$ is neglected, because it is 
vanishingly small within the solar system. 

For the metric parameters $B'(r)/B(r)$ and $A(r)$, the energy density $\rho$ 
and the field $F(R)$, regularity at the origin was separately tested with the 
considered set of equations (as was done in \cite{HMV}).

\begin{table}
\begin{tabular}{|c |c |} 
\hline 
Mission &Measured PPN-parameter constraints \\ 
\hline 
Cassini &$\gamma-1= (2.1\pm2.3)\times 10^{-5}$ \\ 
\hline 
VLBI&$\gamma-1=(-2\pm 4)\times 10^{-4}$ \\ 
\hline 
LLR &$4\beta-\gamma-3=(4.4\pm 4.5)\times 10^{-4}$\\ 
\hline 
Mercury perihelion&$|2\gamma-\beta-1|<3\times 10^{-3}$  \\ 
\hline 
\end{tabular} 
\caption{Observational constraints from solar system missions:  
Cassini spacecraft mission \cite{cassini}, Very Long Baseline  
Interferometry (VLBI) \cite{VLBI}, Lunar Laser Ranging (LLR) tests  
of general relativity \cite{lunar} and the Mercury perihelion shift  
measurements \cite{merc,Will}. The constraints for $\beta_{PPN}$ are not 
discussed in this paper, as the Cassini mission results give the most 
stringent limits for SdS here even for $\beta_{PPN}=1$.}
\label{PPNs} 
\end{table}

\subsection{On polytropic model}\label{sec_polytr} 

We use standard polytropic configurations  
that resemble  solar mass main sequence stars that are like the young sun. 
We chose a polytrope with solar mass and radius as the reference star 
with the polytropic index $n_p=3$. 
This is the socalled Eddington model, and it is given by
 the general relativistic Tolman-Oppenheimer-Volkoff (TOV) equation
in the non-relativistic limit.

As selfgravitating globes of gas, the stars are kept in hydrodynamic  
equilibrium with the  equilibrium condition 
\be{hydroeql} 
\dd p {x^{\alpha}} = \left( \rho+ \f{p}{c^2}\right) F_{\alpha}.
\ee 
Here $F_{\alpha}$ is the gravitational force \cite{WEI}.
By using the general spherically symmetric metric (\ref{metric}) and static, 
distance dependent perfect fluid
\be{p-fluid}
T_{\mu\nu}=p(r) g_{\mu\nu} + (p(r)+\rho(r))\mathrm{u}_\mu \mathrm{u}_\nu 
\ee
(here $\mathrm{u}_\mu$ represents the 4-velocity of the fluid),
the Ricci tensor component $R_{00}$ can be rewritten with the 
hydrostatic equilibrium condition to yield the TOV equation:

\bea{tov}
-r^2 p'(r)&=&G \mathcal{M}(r) \rho(r) \left[1+\f{p(r)}{\rho(r)}\right] \nonumber\\
&\times&\left[1+\f{4\pi r^3 p(r)}{\mathcal{M}(r)}\right]
\left[1-\f{2 G \mathcal{M}(r)}{r}\right]^{-1}.
\eea
Here $\mathcal{M}(r)$ is the integrated mass of the object
\be{totmass}
\mathcal{M}(r)=4 \pi\int_0^r \rho(r) r^2 dr.
\ee

The polytropic equation of state is parametrized for these static stellar objects 
as
\bea{polytr} 
p&=&K\rho^{\gamma}.
\eea 
Parameter $\gamma=1+1/n_p$ is usually written with the
polytropic index. The value $n_p=3$ corresponds to the Eddington polytropic solution  
for the Lane-Emden (LE)  
equations, where solar mass $M_{\odot}$ and radius $R_{\odot}$ are input into the model.
This gives a reasonable fit  to the standard solar model \cite{ssm}, which well reproduces 
the luminosity and neutrino output as well as helioseismological observations.

According to \cite{polystab}, the solar polytropic model with index  
$n_p=3$ is not stable  with respect to Jacobi stability analysis.
The solar polytropic index is, however, stable within linear stability  
analysis and Kosambi-Cartan-Chern theory (see \cite{polystab} and references 
therein). 
The polytropic index values: 
$(11+8\sqrt{2})/7 \leq n_p \leq 5$ 
would conform to  
also Jacobi stable polytropes. However, when departing from the solar index,
the radius-mass relation changes dramatically. The behaviour of the  
polytropic star with the used field equations would also change for this index,
and new degrees of freedom would be needed in the analysis.
Therefore, we perform our analysis with the standard Eddington polytropic index. 

The interconnected Lane-Emden coefficients 
($K$, $\gamma$ in (\ref{polytr}) and scale length $l$ in LE equations  
\cite{WEI})  need to be chosen accurately in order to obtain a star that best 
describes the  sun.
This occurs with the standardized Eddington model
with the solar central density 
\be{rhoinit}
\rho_i=\rho_\odot=76.5 \textrm{ g cm}^{-3}.
\ee 

We parameterized our polytropes (\ref{polytr}) as when deriving
the Lane-Emden equations: 
\be{LEprmz} 
\rho(r) = \rho_{i} \theta(r)^{1/(\gamma-1)}.
\ee 
With this setup, the solar polytrope has $n_p=3$, $\theta_0=1$ and (\ref{rhoinit}).
In the numerical study,
we also varied the central density of the Eddington polytrope in order to find 
statistically significant differences or similarities between 
the modified gravity polytropes and general relativistic ones. 

Also, other realistic main sequence stars ($M_{HS}\neq M_{\odot}$)  
were searched from larger polytropic parameter space with this $f(R)$ gravity.  
But as in the case of general relativistic $n_p=3$  polytropes, 
the only possible configurations are those with the solar mass. 
This occurs in GR
 because $M\sim R^{(3\gamma -4)/(\gamma -2)}$ 
(see \eg (11.3.18) in \cite{WEI}).
The mass $M$ is constant for the Eddington model $n_p=3$.

The \lcdm model was produced by integrating the equation ansatz
with the function $f(R)=R+\Lambda$  and Eddington polytrope parameters.
Here $\Lambda$ is provided by
the cosmological background density $\rho_{\Lambda}$ used in (\ref{vac}). 
This comparison model provides the reference value for $(\gamma_{PPN}-1)$.
Within the considered numerical accuracy, the high curvature regions 
the \lcdm model refers to general relativity (\ie $f(R)=R$).
Therefore, the reference
model is here denoted the \lcdm model.

\section{Applications}\label{numerical} 

We compared  the Hu-Sawicki $f(R)$ model to the \lcdm model 
locally to find out if 
differences between GR and different HS models could be distinguished.
We studied  the  Hu-Sawicki $f(R)$ gravity inside and at the surface  by   
solving for a sunlike matter distribution $\rho(r)$ from the center of the  
sphere to the boundary and calculated the \gppn parameter at the boundary to be able to compare the
polytropes to the observations.

The thin-shell mechanism is not  
discussed in this paper, because all the conclusions here  
are made either inside  the polytrope or at the surface by direct numerical calculations.
We focus on the 
high/low curvature behavior of this $f(R)$ model inside the matter sphere and
on the boundary conditions
without resorting to the scalar-tensor equivalence.

For the used equation ansatz (in \ref{eqns}) the integration variables inside 
the polytrope are the metric components
$A(r)$ and $B(r)$, scalar curvature $R(r)$ and the energy density parameter 
$\theta(\rho(r))$ form (\ref{LEprmz}).

We define the center within the numerical integration
to be under one meter. 
SI units with $c=1$ were used throughout this study. 
The spacetime around the polytrope was found by integrating from the
central region outwards 
until a boundary is found at   
$\rho(r)=10^{-5}$  kg m$^{-3}$, as the sun's photosphere  lies at  
$\rho_{ps}=2\times 10^{-4}$ kg m$^{-3}$. 
In a polytropic configuration, the energy density always becomes negative 
at some point, and this determines the surface.
This definition, therefore, gives a discontinuous surface for the polytropes,
and the SdS vacuum is glued on top of the solution at the boundary.
The TOV equations always arrive at $(\gamma_{PPN} -1)\simeq 8\times10^{-7}$ 
at the stellar boundary. This is  the
expected reference GR value that is compared to the spacetime parameter of 
$f(R)$ gravity polytropes.

\subsection{Constraining the parameter space}\label{prmconstr} 
To get a grip of the dynamics of this non-linear system with several free 
parameters and four initial parameters, 
one needs to narrow down the parameter space.

First we constrained the parameter space according to physical conditions
and then scanned a wide range to find out how these gravity models behave
around a polytropic sphere.

The initial values 
$A_0$, $B_0$ for the spherically symmetric metric 
coefficients (\ref{metric}) were set by using the regularity condition  
at the center.
We, therefore, required the equation ansatz to be regular as $r\rightarrow 0$
with linear, second order or third order perturbations.
Note here that the coordinates $\Theta$ and $\Phi$ in (\ref{metric}) 
do not enter the used equations at all.

For studying the possible spacetimes around the $f(R)$ polytropes, 
we  scanned a large parameter space of $n,c_1,c_2,R_0,\theta_0$ 
(of which $R_0$ and $\theta_0$ are initial values at the center). 
The scalar curvature initial value $R_0$ was scanned from the 
vacuum value $10^{-35}$
to the solar central value $\kappa \rho_\odot$. 
This parameter effectively determines what the external curvature 
will be at the boundary.

We constrained the function (\ref{fHS}) 
by considering only those parameters $c_1$ and $c_2$
that give the correct static vacuum value for the Ricci scalar $R_{VAC}$.
This was carried out by first solving for the scalar curvature from the 
trace equation (\ref{traceq}) with the vacuum energy density
\be{vac} 
T_{\mn}=\kappa^2 \rho_{\Lambda}
\ee 
and $\rho_{\Lambda}$ of the order $10^{-27}$ kg $\textrm{m}^{-3}$ with $c=1$.
The correct branch for these solutions has a real and positive $R$, and equate
to $R_{VAC}=12 H_0^2$
(where $H_0$ is the Hubble value today).  
Due to this requirement, the Hu-Sawicki parameters 
$c_1$ and $c_2$  have a fixed ratio $c_1/c_2=10$. 
Also, in the original paper
\cite{HS}, a similar ratio was obtained by defining
\be{c1c2}
\f{c_1}{c_2}\approx 6 \f{\Omega_{\Lambda}}{\Omega_m}.
\ee
This is of the same order of magnitude as our ratio derived by the procedure above.
The ratio did not, however, have the effect on the exterior spacetime. The \gppn parameter
was always near 1, even if the parameters $c_1$ and $c_2$ differ considerably.

The free model parameters $n$ and $c_2$ were scanned over values $n\in[1,12]$,  
$c_2\in[10^{-10}, 10^{10}]$ to study more widely the model behavior.
And since $f_{R0}=-n c_1/c_2^2/(41)^{n+1}$ this range covers the values
$f_{R0}\in [0.1,10^{-7}]$  
that were also used in the Hu and Sawicki paper \cite{HS}.  The values of $c_2$ 
that do not fulfill the above condition for $f_{R0}$ were treated separately.
The $f(R)$ function with the considered parameter values also satisfied the following conditions
\bea{fRconds}
\f{f'(R)}{f''(R)}& >& 0,\label{sei} \\
f''(R) &> &0, \\ 
f'(R)& \ll& 1, \label{HScond1} \\
\f{f(R)}{R}& \ll& 1.  \label{HScond2} 
\eea
These are the stability condition (\ref{sei}) for a quasi-Newtonian stellar interior 
($R\simeq \rho$) \cite{Seifert} and
the stability condition in high curvature regime \cite{SHS}, where
(\ref{HScond1}) and (\ref{HScond2}) also hold \cite{HS}.

The resulting spacetime parameter \gppn, mass and radius of the polytrope 
seem highly  degenerate over these parameter values. For  the 
HS parameters $n,c_2$ over the studied range, the maximal change in \gppn parameter was 
of the order $10^{-6}$ for all $n$ and $c_2$.
Therefore, for this equation ansatz, the polytrope seems to be effectively 
determined by the polytropic equation of state and the initial scalar curvature.

The polytropic parameters for the solar Eddington model with  $\theta_0=1$ in (\ref{LEprmz})
were discussed in \ref{sec_polytr}.

Finally, in order to compare Hu-Sawicki model to \lcdm \, more statistically, 
we also spanned the initial parameter 
space to include several classes of objects, not just sunlike stars.
The central density parameter $\theta_0$ was also varied
through the values that produce physical central densities 
of red giants $\rho_0/\rho_{0,Edd}\simeq 0.01$ to a very dense star with  
central density  $\rho_0/\rho_{0,Edd}\simeq 1000  $. 
In this respect, this class of models behave very similar to the \lcdm model.
The polytropic energy density parameter at the center, $\theta_0$, is an effective parameter 
wrt the \gppn -parameter with the order of $10^{-5}$ differences at the boundary. 
Like with the initial scalar curvature.
Results of sun like objects (with $\theta_0\sim 1$), that all produce 
\gppn of order 1 are presented in the figures.

To note, Hu-Sawicki model has also been compared in \cite{sneGRB} to 
 Type Ia supernova  (SneIa) and gamma ray burst (GRB) data.
The parameter values favoured by SneIa and  
GRB data with $95\%$ CL in that work do not conform to the correct 
static vacuum solutions  for this class of models.

\subsection{HS polytropes} 

For the scanned range of initial scalar curvature values, the spacetime 
outside the sphere is near to, but not exactly, either 
SdS or Einsteinian (corresponding to $f(R)=R$).
This is easy to see if the function $f(R)$ is written as in (\ref{modfHS}),
where the last term vanishes for all initial values except near the vacuum 
$R\simeq m^2$ 

Scalar curvature $R$ in this system shows stabilising behavior starting in the core area.
The $f(R)$ solution
always approaches to  fulfill the high curvature  solution 
$R(r)\simeq \kappa\rho(r)$ and closely follows this solution throughout the rest of the polytrope.
Because of this tracking behaviour, initial scalar curvature $R_0$ can be almost  
anything and still produce a GR-like solution. 

Furthermore, the high curvature limit for the HS parameter $f_{R0}\ll 1$ did not
correlate with the Ricci scalar behaviour. Effectively all the solutions arrived at
a high curvature solution $R(r)\simeq \kappa \rho(r)$, irrespective of
the value of the $f_{R0}$ parameter.
We also find out that
the calculated Post Newtonian parameter $\gamma_{PPN}$ fulfills the observational 
bounds in less than $40\%$ of the polytropes with varying initial curvature.

\begin{figure}[t] 
\begin{center} 
\includegraphics[width=0.5\textwidth,angle=0]{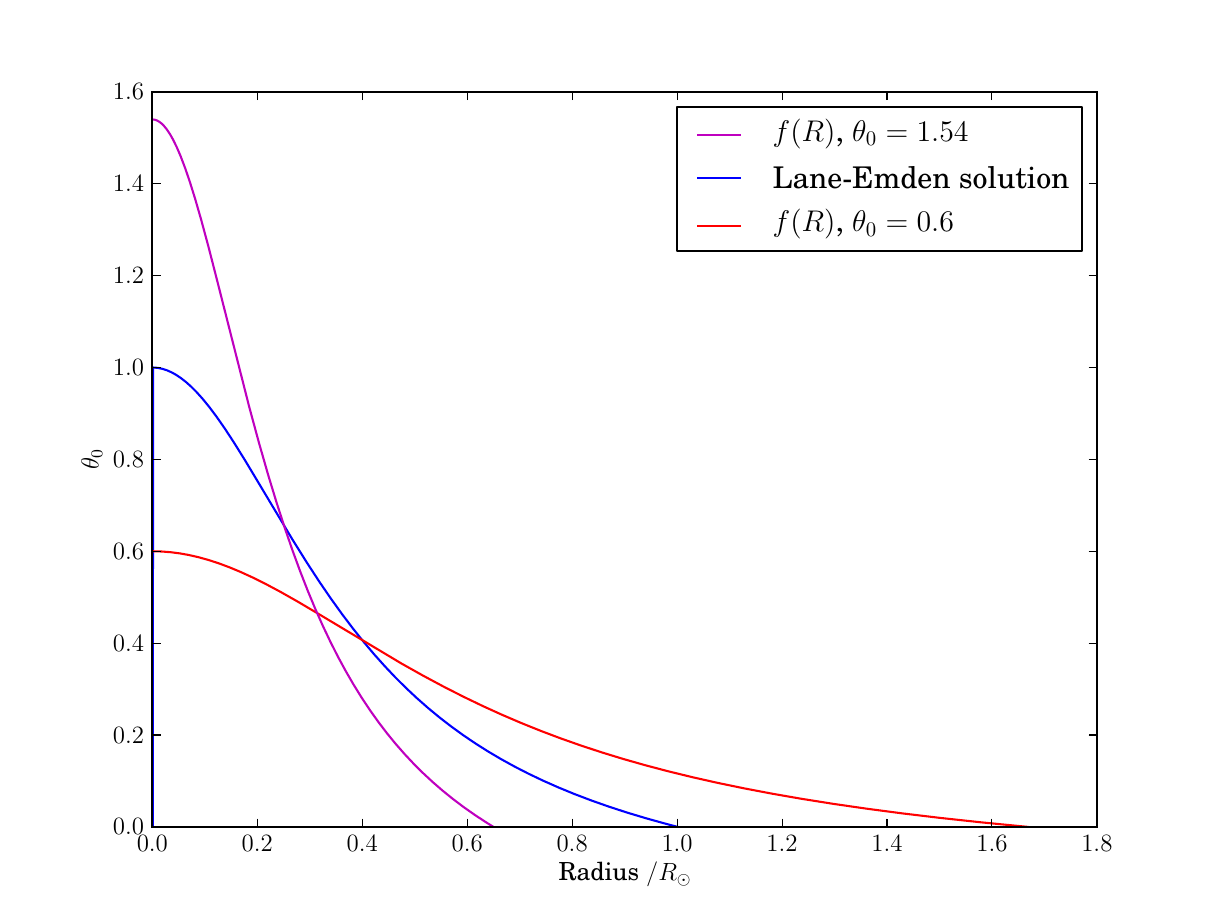} 
\caption{General relativistic LE polytrope is compared to Hu-Sawicki $f(R)$  
polytropes. By using the Lane-Emden equations a variety of stellar types with solar mass
are studied to find out how well HS $f(R)$ models are able to reproduce general relativistic stars. 
The radius of the $f(R)$-polytropes change inversely proportional  
to the central density parameter $\theta_0 $ in (\ref{LEprmz}), 
such that the mass of the polytrope is always the solar mass $M_\odot$.
For the plots in this figure, the central density $R_0$ is constant. 
The Eddington central density gives the solar radius and mass  for  $\theta_0=1$ 
with both gravities, GR and HS.
(Color online)
} 
\label{polytropes} 
\end{center} 
\end{figure}

When considering the Cassini observations, the only effective parameters 
in this setting for distinguishing Hu-Sawicki model from 
the \lcdm  are the central density  $\theta_0$ in (\ref{LEprmz}) 
and the initial curvature $R_0$. 
The two models don't, arrive at the same spacetime at the boundary.
For the central density of the Eddington  
polytrope ($\rho_\odot=76.5$ g cm$^{-3}$ and $\theta_0=1$), the spacetime parameter outside the polytrope  
varies around the SdS-value with a mean value near the \lcdm value everywhere in
the parameter space $\theta_0 \in [0.5, 1.5]$ 
(see FIG.\ref{polytropes}). 

The scatter in \gppn parameter for certain polytropic model 
variates effectively only wrt the initial curvature.
We present the results from a wider polytropic set (not just for
the Eddington polytrope) to be able to statistically compare 
these two  gravity models in FIG.\ref{cassiAc13_mean} and \ref{data_all}. 

The reference value for a general relativistic Eddington polytrope at the 
defined surface approaching from inside is  
$\gamma_{PPN,\Lambda CDM}=0.9999991717$. It is exactly one once the vacuum is  
reached, since the polytropes surface is discontinuous at the border 
by the definition of our polytropes.

\begin{table} 
\begin{tabular}{|c |c c| c c c |} 
\hline 
&\lcdm& & &$f(R)_{HS}$& \\ 
\hline 
$\theta_0$&$R/R_\odot$&$M/M_\odot$ &$R/R_\odot$&$M/M_\odot$&$\gamma_{PPN}-1$ \\ 
\hline 
0.5&2.030&1.0484 & 2.000&1.0025& 0.00053604 \\ 
0.6&1.692&1.0484 & 1.666&1.0025& 0.00001434 \\ 
1.0&1.015&1.0484 & 1.000&1.0024& 0.00040438 \\ 
1.2&0.846&1.0484 & 0.833&1.0024& 0.00233362\\ 
2.0&0.508&1.0484 & 0.500&1.0024& -0.00005374 \\ 
\hline 
\end{tabular} 
\caption{ 
GR values are compared here to modified gravity  
$f(R)_{HS}$-model values for different fiducial central densities $\theta_0$, 
parametrized in (\ref{LEprmz}).  
$f(R)$ polytropes with $\theta_0=0.5$ and $2.0$ 
are also shown in  Fig.(\ref{polytropes}). 
For reference, the spacetime parameter for the \lcdm model is  
(\gppn-1)=$-8\times10^{-7}$ in our work. 
}\label{tab} 
\end{table} 

Because the equations in chapter \ref{eqns} are of higher order than the 
field equations of general relativity, the conditions at the surface of the  
star need to be more stringent. 
Extra gravitational degrees of freedom emerge when the action (\ref{action})
is not linear with respect to the Ricci scalar $R$.
For example, the  metric parameters $\ddot{B}$ and $\dot{A}$ in (\ref{metric})
need not be continuous at the surface of the star in general relativity, 
but in the case of fourth order gravity are required to be continuous. 
Also, the initial value $R(0)$ can be freely chosen here, as the 
equations of motion are of higher order than in general relativity.

Although very similar, the two models, HS and \lcdm, differ systematically within 
observational accuracy.
A key issue that distinguishes GR and $f(R)_{HS}$ polytropes
with this setup is the spacetime at the boundary. 
Although near, the \gppn parameter is not sufficiently close  
to $1$ around the $f(R)$ polytropes for the majority of the solutions
wrt to the Cassini observations. 
According to our calculations, the HS model parameters $n$ and $c_2$ do not effectively 
determine the spacetime outside the polytrope; this is determined by the 
central Ricci scalar value and $\theta_0$. Therefore, the
spacetime outside is determined by the polytropic parameters
and the nonlinearity of the field equations.
This behaviour can be understood by the non-Birkhoffian nature of the system.
Because the system is determined by a fourth order differential equation,
the Birkhoff's theorem is no longer valid \cite{MV}.
Therefore, more initial conditions than the bare stellar mass is needed to determine the solution.
This is true albeit the HS-model resembles the \lcdm model in high curvature:
the higher derivatives do not vanish exactly but maintain the non-Birkhoffian phenomena.

\subsection{Numerical results} 
All the solutions were required to be regular at the origin.  
Even the regularity condition was not sufficient to force the solution  
to be the Schwarzchild-de Sitter spacetime.
Although the solution for the scalar curvature 
$R(r)$ is non-linear near the center, all the initial values
produced similar solutions at the boundary.
Numerical problems occur when the initial scalar curvature is
near the solar central value  $\kappa \rho_\odot$. All the polytropes are not solvable in those cases.

The equations (\ref{eequs},\ref{traceq},\ref{cont})
were solvable for only a small set of initial values when
$n\geq9$ in (\ref{fHS}). This might be due to the
high degree of non-linearity in the equations.
Also, the spacetime parameter \gppn turns out to be far from 1  when $n\geq 9$.

Overall, for $n\in[1,8]$ in (\ref{fHS}) with the chosen ansatz, 
higher than 13 effective digit accuracy was not feasible for a large set of solutions.
This accuracy, however, is sufficient for finding an adequate  $\gamma_{PPN}$ 
value, for the definition (\ref{gamppn}) requires only 7 digit accuracy.
When the accuracy requirement is tightened, the amount of solvable polytropes drops  
dramatically because of numerical difficulties.
However, the mean $\gamma_{PPN}$ value does not converge to the \lcdm value even if 
the accuracy is  increased. The mean is presented in FIG.\ref{data_all}.

Outside the correct parameter area confronting the static vacuum solution for  
the trace equation, $f(R)_{HS}$ polytropes do not yield 
$\gamma_{PPN}=1/2$ outside, as in the case of the simpler polynomial  
$f(R)$ models. 
The $\gamma_{PPN}$ value rather lies near 1 everywhere except if the 
initial value for the scalar curvature is that of the vacuum. 

As a  result, the PPN parameter values  typically  lie within 
$|\gamma_{PPN}-1|<1\times 10^{-4}$, \eg  
the solutions usually tend to a SdS-like spacetime.
SdS spacetime is, however,  rarely reached. 
Typically less than $40\%$ of the solar (\ie $\theta_0\simeq1$)
solutions at the surface
have the SdS spacetime parameter within the 
acceptable observational limits (\ref{cassinig}). 
With only $10\%$, in a typical $29 000$ polytrope run, converging to  
$(\gamma_{PPN}-1)\leq 10^{-6}$. 
With a mean value  
$\left<\gamma_{PPN}-1\right> = -6\times 10^{-6}$. 
The \lcdm value calculated with $f(R)=R+ \Lambda$ in the code gives 
$(\gamma_{PPN,\Lambda CDM}-1)=-8\times 10^{-7}$.

\begin{figure}[th] 
\begin{center} 
\includegraphics[width=0.5\textwidth,angle=0]{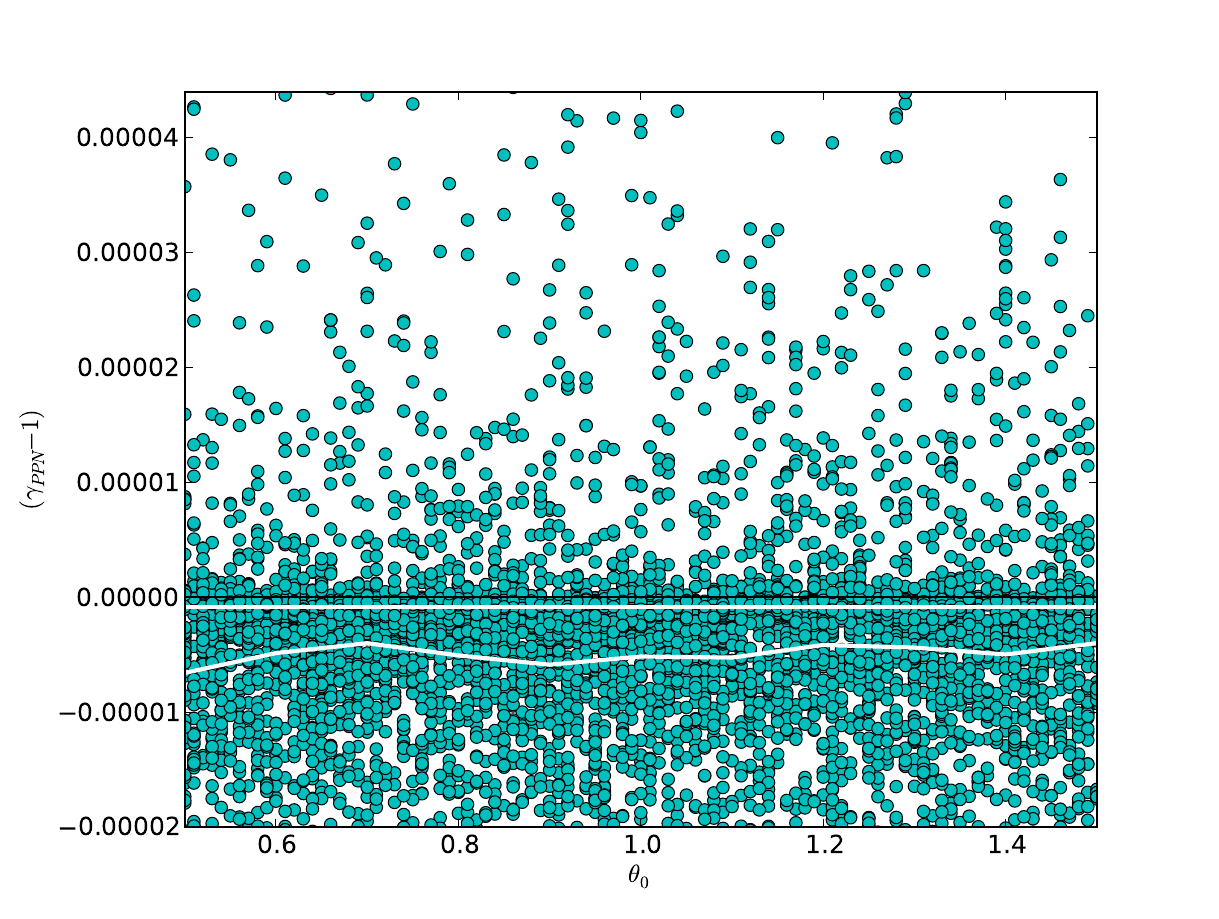} 
\caption{The (\gppn-1) value is plotted with respect to varied central  
energy density parameter $\theta_0$ in (\ref{LEprmz}). 
The scatter in (\gppn-1) comes from the varied values 
of the initial curvature $R(r=0)$. Hu-Sawicki parameter $n=4$ for this plot.
The observational values according to Cassini mission (\ref{cassinig}) reach from
$-2\times 10^{-6}$ to $4.4\times 10^{-5}$.
Lowest curve shows the mean value $<$\gppn$-1>$ per initial central density  
$\theta_0$.
For comparison, a constant line (the upper white line) 
presenting the \lcdm model value is also plotted. (Color online)
}
\label{cassiAc13_mean} 
\end{center} 
\end{figure} 

\begin{figure}[h!] 
\begin{center} 
\includegraphics[width=0.5\textwidth,angle=0]{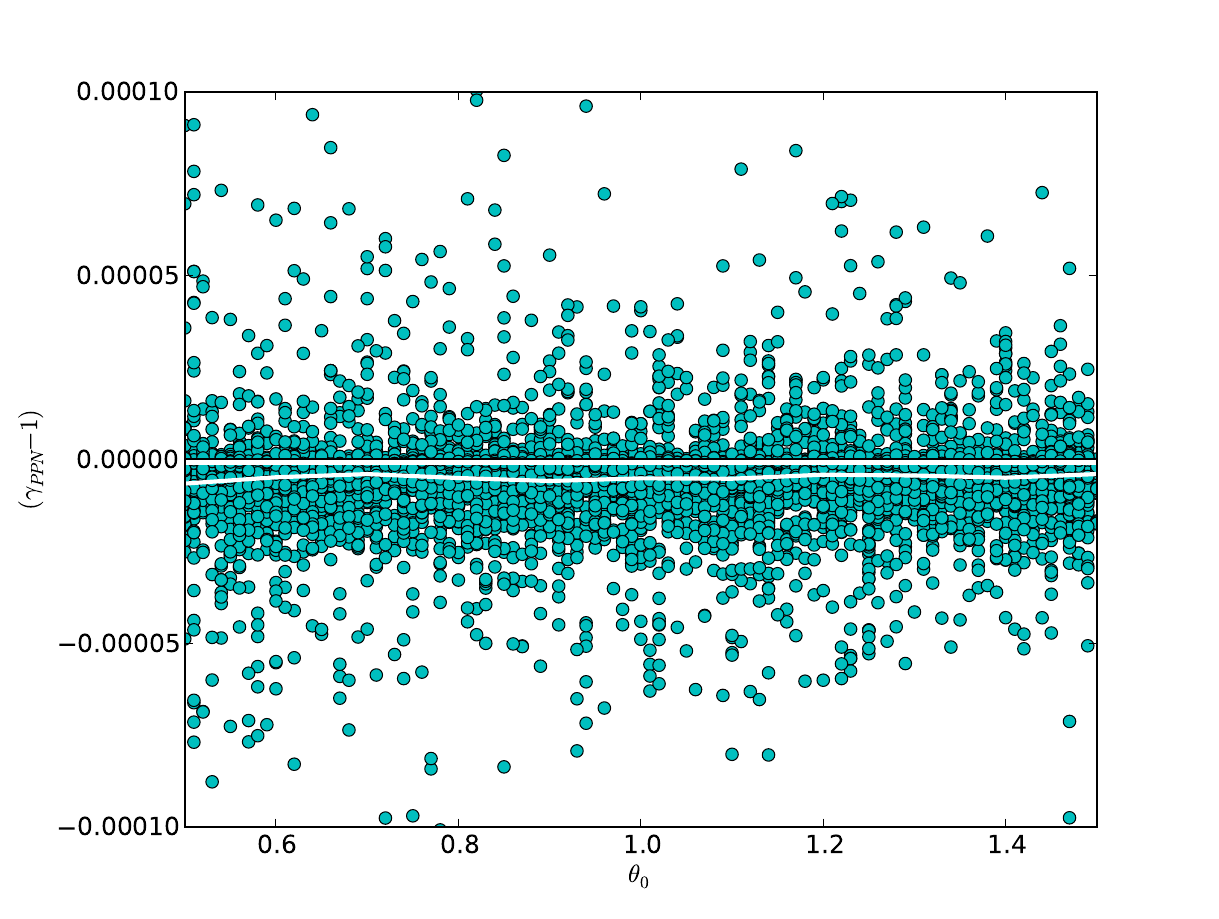} 
\caption{
Approximately one third of the polytropic
solutions lie within the Cassini limits 
(\gppn-1)$= (2.1\pm2.3)\times 10^{-5}$, shown more closely in FIG.\ref{cassiAc13_mean}. 
The data concentrates close around the general relativistic value 
\gppn=1, which is a special feature of this model.
In this plot the HS model $n=4$ is shown,
for $n=1,2$ the mean value $<$\gppn$-1>$  lies marginally closer to the \lcdm line,
still not within the observational bounds. (Color online)
}
\label{data_all} 
\end{center} 
\end{figure} 

In FIG. \ref{data_all} the whole set of polytropes is shown.
In both figures, the \lcdm value is also plotted for reference. 
It is always closer to the SdS value $\gamma_{PPN}=1$, 
than the Hu-Sawicki mean value.
The mean \gppn value for the Hu-Sawicki polytropes converges to a value
that lies within the observational bounds but is always different  
from the \lcdm value. 
In FIG.\ref{cassiAc13_mean}, the  mean values for the spacetime parameter $\gamma_{PPN}$
are calculated over varying initial central density $\theta_0$.  
Here only the values near the observationally acceptable range are plotted. 

We found that the mass of a $f(R)$ polytrope is always closer to the observed  
solar mass than the mass of the Eddingon polytrope. The difference in the masses 
$M_{HS}/M_{\odot}\simeq 1.002$ and  $M_{GR}/M_{\odot}\simeq 1.048$ is systematic 
for all the $n_p=3$ polytropes. 
Although the solutions are similar with respect to the initial values and obtained 
radii and masses, the solutions are never identical.

The $f(R)$-model is stable with respect to short-timescale instabilities 
\cite{SHS} if 
$f''(R)$ is positive. For the present model, this stability condition is satisfied 
with extremely small, positive $f''(R)$ everywhere inside and near the polytrope. 
The mass term  within the corresponding Brans-Dicke 
framework is  dependent on  a $1/f''(R)$ term.
The mass term can be written as in \cite{CSE}
\be{mass2}
m_\sigma^2=\f{(f_0')^2-2 f_0 f_0''}{3 f_0' f_0''},
\ee
evaluated at the constant curvature $R_0$.
Considering the scalar-tensor representation the mass of the scalar, $m_\sigma^2$, is
 in this case always big enough for the ``fifth force'' problem to be evaded
for these polytropic configurations.

\section{Conclusions and discussion}\label{conclu}
Hu-Sawicki polytropic solutions are indeed very similar to general  
relativistic polytropes. The curvature always tracks to the general  
relativistic high curvature solution $R\sim \kappa\rho$ throughout most of the stellar radius, 
and  the radius and mass of the HS-polytrope are 
similar to the GR solution.

The notable difference, however, 
is the spacetime around the configurations. 
In less than $40\%$ the $f(R)_{HS}$ polytropes, in a typical 
29 000 polytrope run, have \gppn parameter 
within the experimental Cassini limits \cite{cassini}.

Many $f(R)$ models result in \gppn$=1/2$ \cite{1perR,CSE,MV}, far from the GR value \gppn$=1$. 
The studied HS model is different in this respect although not fully consistent 
with the observations. 
For the Hu-Sawicki model, $\gamma_{PPN}$ value is nearly one everywhere in our 
model parameter space when $n<9$.
The $n\geq 9$ models had \gppn far from 1 and are not discussed further here.

However, the $f(R)_{HS}$ spacetime around a spherically symmetric  
configuration with a polytropic equation of state nearly always converges 
to a different  
spacetime than is allowed by GR. 
The \gppn -parameter is, however,
always very close to value 1, and no thin-shell mechanism (\ref{HSchapt},\ref{numerical}) is needed to bring
the external spacetime near the observational value.
As the scalar curvature  always stabilizes to the GR value ($\kappa \rho$)
near the center,  no specific initial value can be  
attached to the HS polytrope for any $n$ or $c_2$ that would exactly correspond to the general  
relativistic polytropic sun. Therefore, an observationally acceptable solution is hard to be
justified by general arguments.
A physical super selection rule might, however, exist. 
We do not suggest any mechanism for this, but we would like to note that 
the sun is the only star for which the SdS spacetime is measured to be 
correct with good accuracy. Could it be that other sun like young stars that 
can be modeled with a polytropic equation of state might arrive at a slightly  
different vacuum than SdS? In this case the HS model could indeed describe  
the spacetime around young main sequence stars more accurately.

The only effective parameters wrt \gppn turned out to be the central curvature and initial
energy density, that is, the polytropic model itself. The Hu and Sawicki gravity
does not play a role in determing the spacetime outside by means of the parameters.
The fact that the equations are of fourth order  seems to  cause the exterior spacetime behavior
to generally depart from  Schwarzchildian.
We interpret this behavior as the manifestation of the non-Birkhoffian nature of the model.

With only $40\%$ of the $\gamma_{PPN}$ values reaching the observational limits
with the Hu-Sawicki gravity,
it is obvious that the HS model does not generally reach the Schwarzchild-de Sitter 
solution outside a polytropic stellar configuration when the Cassini observations are considered. 
The polytropes seem to be highly degenerate over the Hu and Sawicki parameters $n$ and $c_2$.
The solutions do not change notably in the mass,  
radius and $\gamma_{PPN}$ when the Hu-Sawicki parameters are varied. 
HS models with $n=1,2$ have
$(\gamma_{PPN}-1)$ values closer to the average (see FIG.\ref{cassiAc13_mean}).
According to this study, $n=1$ model is marginally better than $n=4$.

The non-linearity of the system poses difficulties, and very high accuracy was not feasible. 
Furthermore, even as the numerical accuracy was increased $<\gamma_{PPN,HS}-1>$ 
value does not converge to $<\gamma_{PPN,\Lambda CDM}-1>$.

The short-timescale instabilities are avoided with extremely small and positive
$f''(R)$.
Also, the scalar field  mass in the Brans-Dicke scenario becomes 
large, as the term $m_\sigma$ is proportional to $1/f''(R)$.

With a different polytropic equation of state, some white dwarfs and neutron  
stars can be modeled. In future work we will study relativistic polytropes, 
namely low density white dwarfs and neutron stars. 
Also, luminosity behavior for a star with modified gravity \cite{lumin} will be an 
interesting point for studying.

\acknowledgments 

We thank Christopher Flynn for insightful discussions on polytropes. 
KH is supported by the Wihuri foundation and Turun yliopistosaatio. 



\begin{thebibliography}{X} 

\bibitem{sn1a} A.~G.~Riess {\it et al.},  
Astron.\ J.\  {\bf 116}, 1009 (1998)
[arXiv:astro-ph/9805201];\\  
S.~Perlmutter {\it et al.}, Astrophys.\ J.\  {\bf 517}, 565 (1999)
[arXiv:astro-ph/9812133];\\  
 P.~Astier {\it et al.} [ The SNLS Collaboration ],
 Astron.\ Astrophys.\  {\bf 447}, 31 (2006)  [arXiv:astro-ph/0510447].  

\bibitem{cmb} D.~N.~Spergel {\it et al.} , Astrophys.\ J.\ Suppl.\   
{\bf 148}, 175 (2003)
[arXiv:astro-ph/0302209]; \\
D.~N.~Spergel {\it et al.}, 
 Astrophys.\ J.\ Suppl.\  {\bf 170}, 377 (2007)
 [arXiv:astro-ph/0603449]. 

\bibitem{lcdm}
  V.~Sahni and A.~A.~Starobinsky,
  ``The Case for a positive cosmological Lambda term,''
  Int.\ J.\ Mod.\ Phys.\ D {\bf 9}, 373 (2000)
  [astro-ph/9904398];
  M.~S.~Turner,
  ``The Case for Lambda CDM,''
  astro-ph/9703161.

\bibitem{bao} 
W.~J.~Percival {\it et al.}  [SDSS Collaboration], 
 Mon.\ Not.\ Roy.\ Astron.\ Soc.\  {\bf 401}, 2148 (2010) 
 [arXiv:0907.1660];

 D.~J.~Eisenstein {\it et al.} [ SDSS Collaboration ],
  Astrophys.\ J.\  {\bf 633}, 560-574 (2005)
  [astro-ph/0501171].

\bibitem{isw}
P.~Fosalba and E.~Gaztanaga,
  Mon.\ Not.\ Roy.\ Astron.\ Soc.\  {\bf 350}, L37 (2004)
  [arXiv:astro-ph/0305468];

 N.~Afshordi, Y.~-S.~Loh, M.~A.~Strauss,
  Phys.\ Rev.\  {\bf D69}, 083524 (2004)
  [arXiv:astro-ph/0308260];

 S.~Boughn, R.~Crittenden,
  Nature {\bf 427}, 45-47 (2004)
  [astro-ph/0305001].


\bibitem{coincid}
 I.~Zlatev, L.~-M.~Wang and P.~J.~Steinhardt,
  Phys.\ Rev.\ Lett.\  {\bf 82}, 896 (1999)
  [astro-ph/9807002].

\bibitem{quint} 
  B.~Ratra, P.~J.~E.~Peebles, 
  Phys.\ Rev.\  {\bf D37}, 3406 (1988); 
   R.~R.~Caldwell, R.~Dave, P.~J.~Steinhardt, 
  Phys.\ Rev.\ Lett.\  {\bf 80}, 1582-1585 (1998). 
  [astro-ph/9708069]. 

\bibitem{teves}
 J.~D.~Bekenstein,
  Phys.\ Rev.\  {\bf D70}, 083509 (2004)
  [astro-ph/0403694];\\
 C.~Skordis,
  Class.\ Quant.\ Grav.\  {\bf 26}, 143001 (2009)
  [arXiv:0903.3602].

\bibitem{string-dim}
 S.~Nobbenhuis,
  Found.\ Phys.\  {\bf 36}, 613-680 (2006).
  [gr-qc/0411093].

\bibitem{etg}
 S.~Capozziello, M.~De Laurentis,
  [arXiv:1108.6266].


\bibitem{fR_intro} 
T.~P.~Sotiriou V.~Faraoni,   
Rev.\ Mod.\ Phys.\  {\bf 82}, 451-497 (2010)
[arXiv:0805.1726];\\   
 S.~'i.~Nojiri, S.~D.~Odintsov, 
 [hep-th/0601213];\\
A.~De Felice, S.~Tsujikawa, 
 Living Rev.\ Rel.\  {\bf 13}, 3 (2010)
 [arXiv:1002.4928]. 


\bibitem{tsujikawa} 
S.~Tsujikawa, 
 Lect.\ Notes Phys.\  {\bf 800}, 99 (2010) 
 [arXiv:1101.0191];\\
 B.~Jain, J.~Khoury,
  Annals Phys.\  {\bf 325}, 1479-1516 (2010).
  [arXiv:1004.3294];\\
  T.~Clifton, P.~G.~Ferreira, A.~Padilla, C.~Skordis,
  [arXiv:1106.2476].

\bibitem{GRtests}
 S.~G.~Turyshev,
  Ann.\ Rev.\ Nucl.\ Part.\ Sci.\  {\bf 58}, 207-248 (2008)
  [arXiv:0806.1731].


\bibitem{chameleon} 
 J.~Khoury, A.~Weltman, 
 Phys.\ Rev.\ Lett.\  {\bf 93}, 171104 (2004)
 [astro-ph/0309300]; 

 J.~Khoury, A.~Weltman, 
 Phys.\ Rev.\  {\bf D69}, 044026 (2004) 
 [astro-ph/0309411]. 


\bibitem{MTW}
  C.W. Misner, K.S. Thorne, J.A. Wheeler,
  ``Gravitation'',
  W.H.Freeman and Company, New York (2000)

\bibitem{WEI} 
 S. ~Weinberg,  
 ``Gravitation and Cosmology'', 
 John Wiley \& Sons, 1972 

\bibitem{MV}
 T.~Multamaki and I.~Vilja, 
 Phys.\ Rev.\  D {\bf 74}, 064022 (2006) 
 [arXiv:astro-ph/0606373];\\
 T.~Multamaki and I.~Vilja, 
 Phys.\ Rev.\  D {\bf 76}, 064021 (2007) 
 [arXiv:astro-ph/0612775]. 
 T.~Multamaki and I.~Vilja,
  Phys.\ Lett.\ B {\bf 659}, 843 (2008)
  [arXiv:0709.3422 [astro-ph]].


\bibitem{birk1} 
 V.~Faraoni, 
 Phys.\ Rev.\  D {\bf 81}, 044002 (2010) 
 [arXiv:1001.2287 [gr-qc]]. 



\bibitem{AT_mr} 
 L.~Amendola and S.~Tsujikawa, 
 Phys.\ Lett.\  B {\bf 660}, 125 (2008) 
 [arXiv:0705.0396]. 

\bibitem{LinGuChen} 
W.~-T.~Lin, J.~-A.~Gu, P.~Chen
Int.\ J.\ Mod.\ Phys.\  {\bf D20}, 1357-1362 (2011)
 [arXiv:1009.3488]. 


\bibitem{Seifert} 
 M.~D.~Seifert, 
 Phys.\ Rev.\  D {\bf 76}, 064002 (2007) 
 [arXiv:gr-qc/0703060]. 

\bibitem{dolgov}   
A.~D.~Dolgov and M.~Kawasaki,   
Phys.\ Lett.\ B {\bf 573}, 1 (2003).   
 [astro-ph/0307285]. 

\bibitem{woodard} 
R.~P.~Woodard, 
 Lect.\ Notes Phys.\  {\bf 720}, 403-433 (2007)
 [astro-ph/0601672]. 


\bibitem{clifton} 
 T.~Clifton, P.~G.~Ferreira, A.~Padilla, C.~Skordis, 
 [arXiv:1106.2476]. 

\bibitem{faraoni} 
 V.~Faraoni, 
   [arXiv:0810.2602 [gr-qc]]. 


\bibitem{HS} 
 W.~Hu and I.~Sawicki, 
 Phys.\ Rev.\  D {\bf 76}, 064004 (2007) 
 [arXiv:0705.1158]. 




\bibitem{AB} 
 S.~A.~Appleby and R.~A.~Battye, 
 Phys.\ Lett.\  B {\bf 654}, 7 (2007) 
 [arXiv:0705.3199]. 


\bibitem{St} 
 A.~A.~Starobinsky, 
 JETP Lett.\  {\bf 86}, 157 (2007) 
 [arXiv:0706.2041]. 

\bibitem{NO1}
 S.~'i.~Nojiri, S.~D.~Odintsov,
  Phys.\ Lett.\  {\bf B652}, 343-348 (2007)
  [arXiv:0706.1378];\\
 S.~Nojiri, S.~D.~Odintsov, 
 [arXiv:0801.4843]. 



\bibitem{NO2}
 S.~Nojiri, S.~D.~Odintsov,
  Phys.\ Rev.\  {\bf D77}, 026007 (2008)
  [arXiv:0710.1738].


\bibitem{cassini} 
 B.~Bertotti, L.~Iess and P.~Tortora, 
 Nature {\bf 425}, 374 (2003). 


\bibitem{koivisto}  
T.~Koivisto, 
Class.\ Quant.\ Grav.\  {\bf 23}, 4289 (2006) 
[arXiv:gr-qc/0505128].  

\bibitem{comp-fRs} 
  L.~Sebastiani and S.~Zerbini,
  Eur.\ Phys.\ J.\ C {\bf 71}, 1591 (2011)
  [arXiv:1012.5230 [gr-qc]];
  W.~Hu and I.~Sawicki,
  Phys.\ Rev.\ D {\bf 76}, 104043 (2007)
  [arXiv:0708.1190 [astro-ph]].


\bibitem{Will}
 C.~M.~Will, 
 Living Rev.\ Rel.\  {\bf 9}, 3 (2005) 
 [arXiv:gr-qc/0510072]. 


\bibitem{HMV} 
 K.~Henttunen, T.~Multamaki and I.~Vilja, 
 Phys.\ Rev.\  D {\bf 77}, 024040 (2008) 
 [arXiv:0705.2683]. 


\bibitem{ssm} 
J.~N.~Bahcall, A.~M.~Serenelli and S.~Basu, 
 Astrophys.\ J.\  {\bf 621}, L85 (2005) 
 [arXiv:astro-ph/0412440]; 
  J.~N.~Bahcall, A.~M.~Serenelli and S.~Basu,
  Astrophys.\ J.\ Suppl.\  {\bf 165}, 400 (2006)
  [astro-ph/0511337].


\bibitem{polystab} 
 C.~G.~Boehmer and T.~Harko,
 arXiv:0902.1054 [math-ph]. 



\bibitem{VLBI} 
 S.~S.~Shapiro, J.~L.~Davis, D.~E.~Lebach, J.~S.~Gregory, 
 Phys.\ Rev.\ Lett.\  {\bf 92}, 121101 (2004). 

\bibitem{lunar} 
 J.~G.~Williams, S.~G.~Turyshev, D.~H.~Boggs, 
 Phys.\ Rev.\ Lett.\  {\bf 93}, 261101 (2004)
 [gr-qc/0411113]. 


\bibitem{merc} 
I.~I.~Shapiro \textit{et.al.} in \textit{General Relativity and Gravitation 12}, 
Cambridge University Press, p. 313 (1990) 

\bibitem{SHS} 
 Y.~S.~Song, W.~Hu and I.~Sawicki, 
 Phys.\ Rev.\  D {\bf 75}, 044004 (2007) 
 [arXiv:astro-ph/0610532]. 

\bibitem{sneGRB} 
 V.~F.~Cardone, A.~Diaferio, S.~Camera, 
   [arXiv:0907.4689]. 


\bibitem{CSE}
  T.~Chiba, T.~L.~Smith, A.~L.~Erickcek,
  Phys.\ Rev.\  {\bf D75}, 124014 (2007)
  [astro-ph/0611867]


\bibitem{1perR}
  S.~M.~Carroll, V.~Duvvuri, M.~Trodden, M.~S.~Turner,
  Phys.\ Rev.\  {\bf D70}, 043528 (2004)
  [astro-ph/0306438];\\
 X.~H.~Jin, D.~J.~Liu and X.~Z.~Li,
  arXiv:astro-ph/0610854;\\
 G.~J.~Olmo,
  arXiv:gr-qc/0505135.

\bibitem{lumi}
  A.~-C.~Davis, E.~A.~Lim, J.~Sakstein and D.~Shaw,
  Phys.\ Rev.\ D {\bf 85}, 123006 (2012)
  [arXiv:1102.5278 [astro-ph.CO]].

\bibitem{lumin}
A.~-C.~Davis, E.~A.~Lim, J.~Sakstein and D.~Shaw,
  Phys.\ Rev.\ D {\bf 85}, 123006 (2012)
  [arXiv:1102.5278 [astro-ph.CO]].
\end{thebibliography}
\end{document}